# Co-GRU Enhanced End-to-End Design for Long-haul Coherent Transmission Systems

Jiayu Zheng, Tianhong Zhang, Yu Wenjing, Weiqin Zhou, Chuanchuan Yang and Fan Zhang, *Senior Member*, *IEEE*, *Senior Member*, *OSA*

*Abstract* — In recent years, the end-to-end (E2E) scheme based on deep learning (DL) has been proposed as a potential scheme to jointly optimize the encoder and the decoder parameters of the optical communication system. Compared with conventional deep neural network (DNN) adopted in E2E design, center-oriented Gated Recurrent Unit (Co-GRU) network has the ability to learn and compensate for inter-symbol interference (ISI) with low computation cost while satisfying the gradient backpropagation (BP) condition. In this paper, the Co-GRU structure is adopted for both channel modeling and decoder implementation in E2E design for long-haul coherent wavelength division multiplexing (WDM) transmission systems, which can enhance the performance of general mutual information (GMI) and $Q^2$-factor. For the E2E system with Co-GRU based decoder, the gain of GMI and $Q^2$-factor are respectively improved 0.2 bits/sym and 0.48dB, compared to that of the conventional QAM system, for a 5-channel dual-polarization coherent system transmitting over 960km standard single mode fiber (SSMF). This work paves the way for further study of the application of the Co-GRU structure for both the data-driven channel modeling and the decoder performance improvement in E2E design.

## I. Introduction

Recently, deep learning (DL) techniques adopting artificial neural networks have been increasingly used in the design and optimization of optical communication systems due to their powerful data learning and fitting ability [1-2].

The End-to-End (E2E) design utilizes the neural network structure to realize the encoder as well as the decoder, providing a feasible scheme for holistic optimization of optical communication systems by modeling the whole system as an auto-encoder [4-6]. With the gradient back propagation (BP) method, End-to-End Deep Learning (E2EDL) performs a joint training of the encoder and the decoder [7-8], which can further explore the system capacity.

E2EDL was first proposed and validated for communication systems with additive white Gaussian noise (AWGN) and Rayleigh channel [9-12]. In particular, combining geometric shaping with probabilistic shaping for the AWGN channel model, E2E achieves a further mutual information (MI) and general mutual information (GMI) gain compared with that through either single means of shaping [13].

For optical communication, the E2EDL scheme was first proposed for intensity modulation/direct detection (IM/DD) systems only considering fiber dispersion induced distortions [14]. Through ELEDL, the bit-to-waveform mapping at the encoder and the wave-to-bit decoding at the decoder are jointly trained to lower the bit error rate (BER) of the IM/DD system [14-15].

For implementing E2E joint training, considering that gradient BP is mathematically rigorous, it is most widely adopted in the E2E design [7]. Therefore, the focus of E2EDL is to select or design a differentiable and computationally tractable channel model to meet the gradient BP condition [8], while keeping the model close to the practical channel. For long-haul wavelength division multiplexing (WDM) coherent systems, the split step Fourier method (SSFM) method provides a feasible channel model close to the real physical channel. Adopting the framework of Tensorflow, the SSFM channel can be realized to satisfy the gradient BP condition. Based on the SSFM channel, E2EDL is applied to the encoder and the decoder parameter joint optimization for nonlinear frequency division multiplexing (NFDM) system, thus achieving a longer distance compared with the manually-optimized system [16]. However, the cost of computing resources, such as Graphic Processing Unit (GPU) memory and running time for forward and BP through the SSFM channel, limits its universal application in E2EDL [7].

There have been several kinds of channel modeling schemes to replace the SSFM channel aiming at allowing gradient BP with lower computation cost. For long haul coherent transmission system performance improvement, there has been several digital signal processing (DSP) algorithms for compensating channel nonlinear distortion [17, 18]. When applying E2EDL instead of traditional DSP algorithms, nonlinear distortion must be considered in channel modeling. For WDM systems, channel modeling based on nonlinear

This work was supported by National Natural Science Foundation of China (62271010 and U21A20454), the major key project of Peng Cheng Laboratory. This work was also supported by the ZTE Industry-University-Institute Cooperation Funds under Grant No.HC-CN-20220616009. (Jiayu Zheng and Tianhong Zhang contributed equally to this work, Corresponding author: Fan Zhang.)

Jiayu Zheng, Tianhong Zhang, Chuanchuan Yang and Fan Zhang are with the State Key Laboratory of Advanced Optical Communication System and Networks, Frontiers Science Center for Nano-optoelectronics, Department of Electronics, Peking University, Beijing 100871, China (e-mail: 2001111276@pku.edu.cn; 2201111564@pku.edu.cn; Chuanchuanyang@pku.edu.cn; fzhang@pku.edu.cn ). Fan Zhang is also with Peng Cheng Laboratory, Shenzhen 518055, China.

Yu Wenjing and Weiqin Zhou are with ZTE Corporation, Shenzhen 518057, China (e-mail: yu.wenjing1@zte.com.cn zhou.weiqin@zte.com.cn).

interference noise (NLIN) model has been applied in E2EDL with geometric shaping, which can achieve further GMI gain over the conventional QAM system [19-24]. The channel adopting NLIN model satisfies the gradient BP condition due to its simplified assumption for the channel nonlinear distortions [19-20]. However, this also leads to a non-negligible gap between the channel model adopting the NLIN model and the real physical channel, thus limiting the equalization ability for practical channel distortions when E2EDL applied based on NLIN. In addition to NLIN, a modeling scheme for the coherent channel adopting the first-order perturbation method is proposed for the memory-aware E2EDL design [24]. However, the bitwise mutual information gain is verified for the single channel system with only 170 km transmission distance [24].

To achieve high modeling accuracy with lower computation complexity compared to SSFM, the data-driven neural network model can be adopted to replace SSFM for channel modeling. Through E2EDL based on data-driven channel model, the IM/DD system can achieve a better bit error rate (BER) performance over that using conventional equalizer [25]. As for coherent systems, replacing the SSFM channel model with deep neural network (DNN) as a differentiable surrogate channel (DSC), the gradient BP condition is satisfied [26]. Testing the encoder and the decoder trained on the SSFM channel, the system gain proves the effectiveness of the E2E scheme that adopts DNN as the DSC for long-haul coherent dual-polarization systems. Meanwhile, the limitation is that the DNN structure is incapable in thoroughly learning and compensating for the channel distortion due to the ISI.

It should be noted that neural network has been used to model fiber link or compensate for fiber nonlinearity in optical communication systems. In particular, recurrent neural networks (RNN) with its related structures such as long short-term memory network (LSTM) and gate recurrent unit (GRU), has been effectively used to simulate optical fiber transmission or equalize fiber nonlinearity distortions [27, 28].

The conventional RNN utilize the information of adjacent symbols to learn and equalize distortions, which has high computation complexity, limiting its application in E2EDL. Recently, we proposed a simplified LSTM structure named as center-oriented LSTM (Co-LSTM) [29], which can effectively equalize fiber nonlinearity with ultra-low complexity. As LSTM can be further simplified to GRU, we can thus extend Co-LSTM to Co-GRU, which provides a feasible scheme for channel modeling and ISI equalization in E2EDL.

In this paper we use ultra-low complexity Co-GRU for channel modeling and decoder training. The E2EDL framework composes of three Phases. In Phase I, the auto-encoder are trained firstly through the NLIN channel model. In Phase II, the decoder and the Co-GRU channel is trained using the data generated by the SSFM model. In Phase III, the encoder is trained through the Co-GRU channel model, while the channel and decoder are updated using the SSFM model. In Phase III, the decoder-only and encoder-decoder joint-training are firstly alternately implemented, then only the decoder is further trained with more epochs. The performance of E2EDL with both the DNN and Co-GRU based decoder are tested on the SSFM channel. E2EDL adopting the Co-GRU based decoder achieves the $Q^2$-factor gain up to 0.48dB and the GMI gain up to 0.2bits/sym compared to conventional 64-ary quadrature amplitude modulation (64-QAM) modulation format for a 32Gbaud 960km 5-channel dual-polarization WDM system.

The rest of this paper is organized as follows. Section II reviews the calculation principle of Co-GRU neural network structure and introduces system modeling with Co-GRU. In Section III, we discuss the end-to-end training process, including the initialization process using the NLIN model. The main results of E2EDL are reported in Section IV. Finally, Section V summarizes the paper and gives the conclusions.

## II. Co-GRU neural network

This section introduces the structure of Co-GRU network and its advantages for channel modeling and equalization.

### A. Co-GRU neural network calculation process

The ability of RNN and its variants such as LSTM network in channel modeling and equalization has been verified. For example, using Bi-LSTM for channel modeling can greatly reduce the computational complexity of forward transmission of channel model. However, the channel model used in E2EDL should meet the requirement of the accuracy and low complexity, and at the same time, the calculation time of its gradient BP should be reduced as much as possible. At the same time, the model used for decoder network and system network should not occupy too much memory resources, otherwise the training will be difficult. Using GPU to calculate the channel with the SSFM model will consume too much computing resources. Bi-LSTM neural network provides a feasible scheme for fiber channel modeling, which requires less computation time than SSFM. However, the calculation time of gradient reverse transmission of Bi-LSTM is too high for E2EDL. Co-LSTM can compensate for nonlinearity in optical fiber communication systems, and its computational complexity is significantly lower than that of Bi-LSTM network [26]. This center-oriented computing scheme can also be applied to fiber channel modeling, which can reduce the computing time of gradient BP and occupy much fewer computing resources compared with Bi-LSTM. Similar to LSTM, GRU network is another variant of recurrent neural network. Compared with LSTM, GRU has fewer parameters and faster convergence. Thus we extend the calculation model of Co-LSTM to the GRU network, namely Co-GRU. This Co-GRU network will be used for the modeling of the decoder network and the part between the encoder and the decoder in optical fiber communication systems.

Assuming that $N$ symbols to be calculated by neural network, and the length of ISI to be considered in channel modeling or equalization is $L$. If the traditional Bi-GRU is used for calculation, $2L+1$ adjacent symbols are required as the input of the neural network when calculating the output at each symbol position, and then the results of the hidden layer will be transmitted to the full connection layer (FCL) to get the final outputs. When using the Co-GRU calculation mode, the main feature is that it does not need to perform calculation for each output symbol. In contrast, we input the entire symbol sequence into Co-GRU and get the result of the hidden layer of each symbol position at one time, and then get the final result by calculating the FCL. We briefly discuss the principle of Co-GRU in the following. Further details can be found in our previous work that introduced Co-LSTM for the first time [26].

$$z_t = \sigma(W_z x_t + U_z h_{t-1} + b_z) \quad (1)$$

$$r_t = \sigma(W_r x_t + U_r h_{t-1} + b_r) \quad (2)$$

$$\tilde{h}_t = \tanh(W_h x_t + U_h (r_t \odot h_{t-1}) + b_h) \quad (3)$$

$$h_t = z_t \odot h_{t-1} + (1 - z_t) \odot \tilde{h}_t \quad (4)$$

Equations (1) - (4) represent the operational structure of a GRU unit. At the moment of $t$, the GRU is composed of an input vector $x_t$, input hidden state $h_{t-1}$ from the moment $t-1$, output hidden state $h_t$, temporary hidden state $\tilde{h}_t$, reset gate output $r_t$ and update gate output $z_t$. The $W$, $U$ and $b$ with the corresponding subscripts represent the weight or bias parameters of different parts, while $\sigma$ represents the sigmoid activation function.

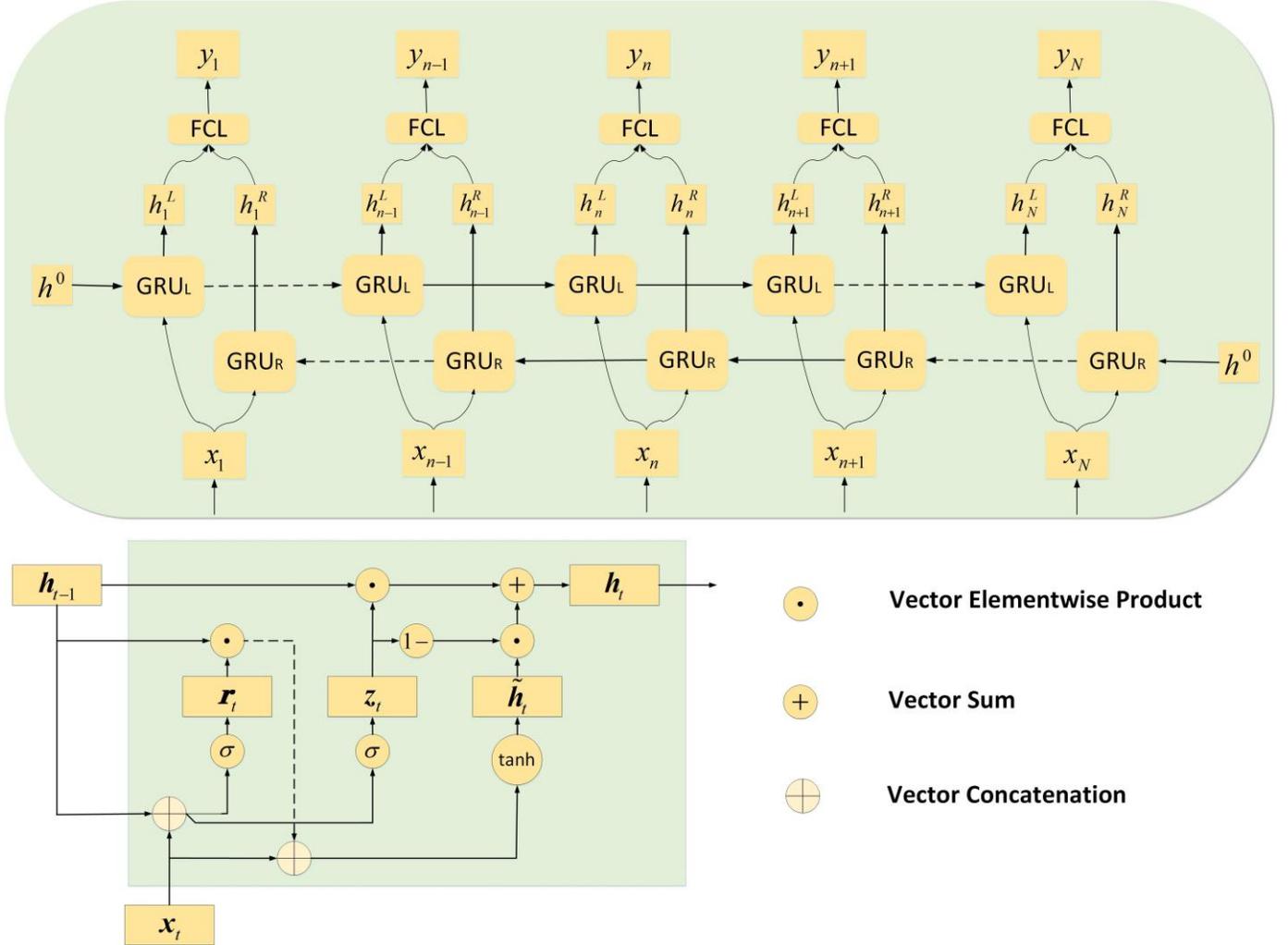

Fig.1. Co-GRU neural network calculation process.(a) The calculating mode of Co-GRU; (b) the structure of the GRU cell.

The schematic diagram of the Co-GRU network used here is shown in Fig.1. $x_o, ... x_{n-1}, x_n, x_{n+1}, ... x_N$ represent the sequence of symbols input, and $y_0, ... y_{n-1}, y_n, y_{n+1}, ... y_N$ represents the sequence of symbols output. For modeling optical fiber communication systems, $x_n$ represents the modulated symbols, and $y_n$ represent the symbols received by the receiver and processed by traditional DSPs. When GRU is used for modeling the decoder at the receiver, $x_n$ represents the symbols that need to be further processed by the decoder, and $y_n$ represents the bit sequence decoded by the decoder.

$GRU_L$ and $GRU_R$ are two GRU units, which handle data transferred to the right and data transferred to the left, respectively. The internal structure of the GRU unit is omitted here. $h_i^L$ and $h_i^R$ represent the result of hidden layer obtained by $GRU_L$ and $GRU_R$ at the $i_{th}$ symbol position. $h^0$ is a zero vector, which is the input as the initial hidden layer.

When calculating the corresponding output at the $n_{th}$ symbol position, the hidden state $h_i^L$ is calculated with Eq.(5).

$$h_n^L \approx GRU_L\left(h_{n-1}^L, x_n, 1\right) \tag{5}$$

At the $(n+1)_{th}$ symbol position, $h_{n+1}^L$ is calculated with Eq.(6).

$$h_{n+1}^L \approx GRU_L\left(h_n, x_{n+1}, 1\right) \tag{6}$$

Therefore, the output of the corresponding hidden layer at each symbol position only needs to be calculated once. When calculating the output of the hidden layer at the next position, the output of the hidden layer at the current position is required as the input of the GRU unit. In doing so, the number of iterations for the whole sequence can be greatly reduced when calculating the neural network. The Co-GRU calculating mode consists of two stages. In the first stage, we use $GRU_L$ and $GRU_R$ to calculate the hidden states $h^L$ and $h^R$ at each symbol position. In the second stage, the hidden states $h^L$ and $h^R$ at the same symbol position are concatenated and then input to the FCL to obtain the output. It should be noted that after the Co-GRU calculation of the symbol sequence, a certain number of symbols with large error in the edge position of the sequence need to be discarded.

In the whole end-to-end training system, we use Co-GRU network for both channel modeling and receiver decoding.

*B. System modeling based on Co-GRU*

The setup of the communication system and the alternative scheme of Co-GRU modeling are shown in Fig.2. Co-GRU network structure is used to model the part between the encoder and the decoder to accelerate the training speed of E2EDL.

In this work, we simulate a 5-channel WDM system with the central channel as the target one. 32Gbaud 64-QAM modulation is applied each channel with a channel frequency spacing of 40GHz. The bit sequence is first mapped to constellation points in the encoder. The coded symbols first go through normalization with the oversampling factor of 16 samples per symbol (SPS). The transmitted samples are reshaped using a root raised cosine (RRC) filter with a roll-off factor of 0.01. The WDM signals are normalized to a specific input power and then enters the channel model. The SSFM channel model is implemented based on Manakov equation. We consider cascaded fiber spans with 80km SSMF each span and lumped erbium-doped fiber amplifier (EDFA) with a noise figure of 5dB. After wavelength demultiplexing, a series of DSPs are applied after coherent detection. The chromatic dispersion compensation (CDC) is first applied. After the matched filter, synchronization, and down-sampling to 2 SPS, time-domain equalization is performed to demultiplex two polarizations. After down-sampling to 1 SPS, the carrier phase recovery is conducted by averaged phase rotation with pilot symbols. Then the signal symbols are sent to the decoder to retrieve the corresponding bits.

We define the part between the encoder and the decoder in this WDM system as the original model. Considering that it is difficult to calculate each DSP step when performing gradient reverse transmission, we use a single Co-GRU network to replace the original model for E2EDL. Co-GRU can model channel distortions and make it possible to mitigate channel distortions by E2EDL.

We compared the calculation time of Co-GRU model, Bi-GRU model and the original model. The Co-GRU model, Bi-GRU model, and SSFM all run on NVIDIA TITAN V Computer Graphics Cards, while the rest of the original model runs on the Intel Xeon CPU E5-2065 v4. We compare the time cost for the transmission of $2^{14}$ symbols using different models. The training cost advantage of Co-GRU network can be seen from Figure 3. Fig.3(a) shows the time required to model the system using Co-GRU, Bi-GRU and the original model. It can be found that for the transmission system with 960km SSMF, the calculation time of the Co-GRU model is 88.7% less than that of the Bi-GRU model and 98% less than that of the original model. The accuracy and feasibility of Co-GRU network can be verified from the loss and the training results in Section IV.

Fig.3(b) shows the comparison of the gradient BP calculation time of different transmission distance. Here we compare the Co-GRU model, Bi-GRU model and SSFM channel calculated by GPU. For the transmission distance of 960km, the calculation time of Co-GRU model is 88.4% less than that of Bi-GRU and 86.8% less than that of SSFM calculated by GPU. It can be found that the Co-GRU model has significant advantages in computing speed of gradient BP. Moreover, when using the Co-GRU model for BP, the calculation time does not increase with the transmission distance. SSFM calculating with GPU has a speed close to Bi-GRU, which, however, takes up too much memory resources for gradient BP as shown in Fig.3 (c).

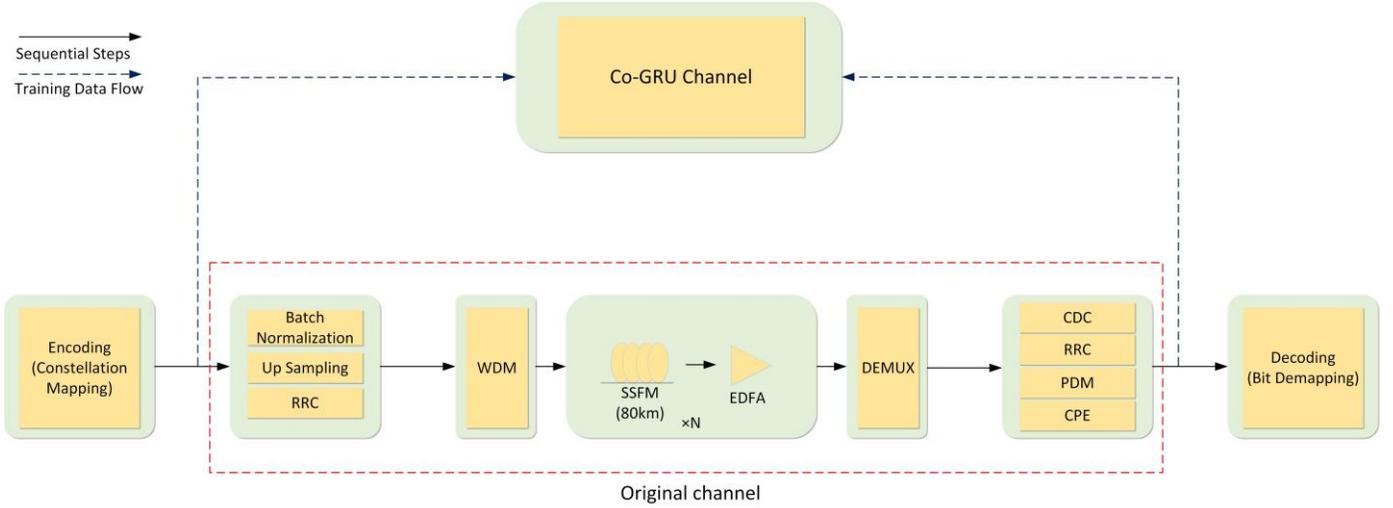

Fig.2. Training of the Co-GRU channel model for replacing the original Channel.

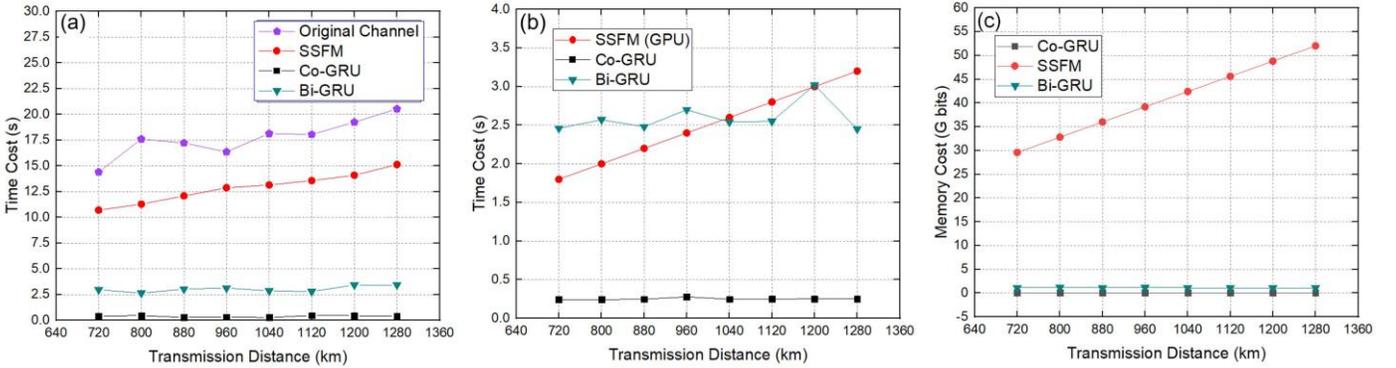

Fig.3. The computation cost of different channel models
(a) The forward transmission time cost of different channel model versus the transmission distance;
(b) The gradient BP time cost of different channel model versus the transmission distance;
(c) The memory cost of SSFM channel and Co-GRU versus the transmission distance

As shown in Figure 4, the E2E training scheme we proposed consists of three Phases (I, II & III).

In Phase I, E2E joint training of the encoder and the decoder is based on the channel with nonlinear distortion simulated by NLIN model. In doing so, we obtain the encoder generating constellation with symmetrical distribution characteristics.

Note that in Phase I, both the encoder and the decoder parameters are initialized by the Kaiming initialization method [30]. This ensures that the constellation generated by the encoder has a random distribution.

In Phase II, the original model with SSFM channel is replaced by Co-GRU introduced in Section II. With the encoder parameters fixed, the Co-GRU channel model and the decoder are respectively trained. For each epoch, the operation ⓑ and ⓒ are respectively implemented 10 times.

In Phase III, the further E2EDL for the encoder and the decoder is implemented N×100 epochs as shown in Fig.4 (b). In contrast to the decoder optimization based on the SSFM channel, the encoder is trained through gradient BP with the Co-GRU channel. Meanwhile, the Co-GRU channel is continually updated to track the geometric shaping of the constellation from the encoder.

We take 100 training epochs as a period as shown in Fig. 4(b). After 25-epoch joint training for the encoder and the decoder parameters, 75-epoch decoder-only training is implemented subsequently. In the first 25 epochs, the operation ⓐ, ⓑ, ⓒ are respectively implemented 2,10 and 2 times for each epoch. While in the last 75 epochs, only the operation ⓑ and ⓒ are both implemented 10 times each epoch.

For each epoch, a new symbol sequence is generated and transmitted. After E2E training in Phase III, the decoder BER almost no longer decreases when the suitable period number selected.

In Phase I, both the encoder and the decoder are implemented with DNN. For Phase II and III, the encoder still adopts DNN, while the decoder is successively implemented with DNN and Co-GRU. The DNN decoder in Phase II is initialized using the result obtained in Phase I, while all the networks in Phase III are initialized using the results trained in Phase II.

## III. END-TO-END TRAINING PROCESS

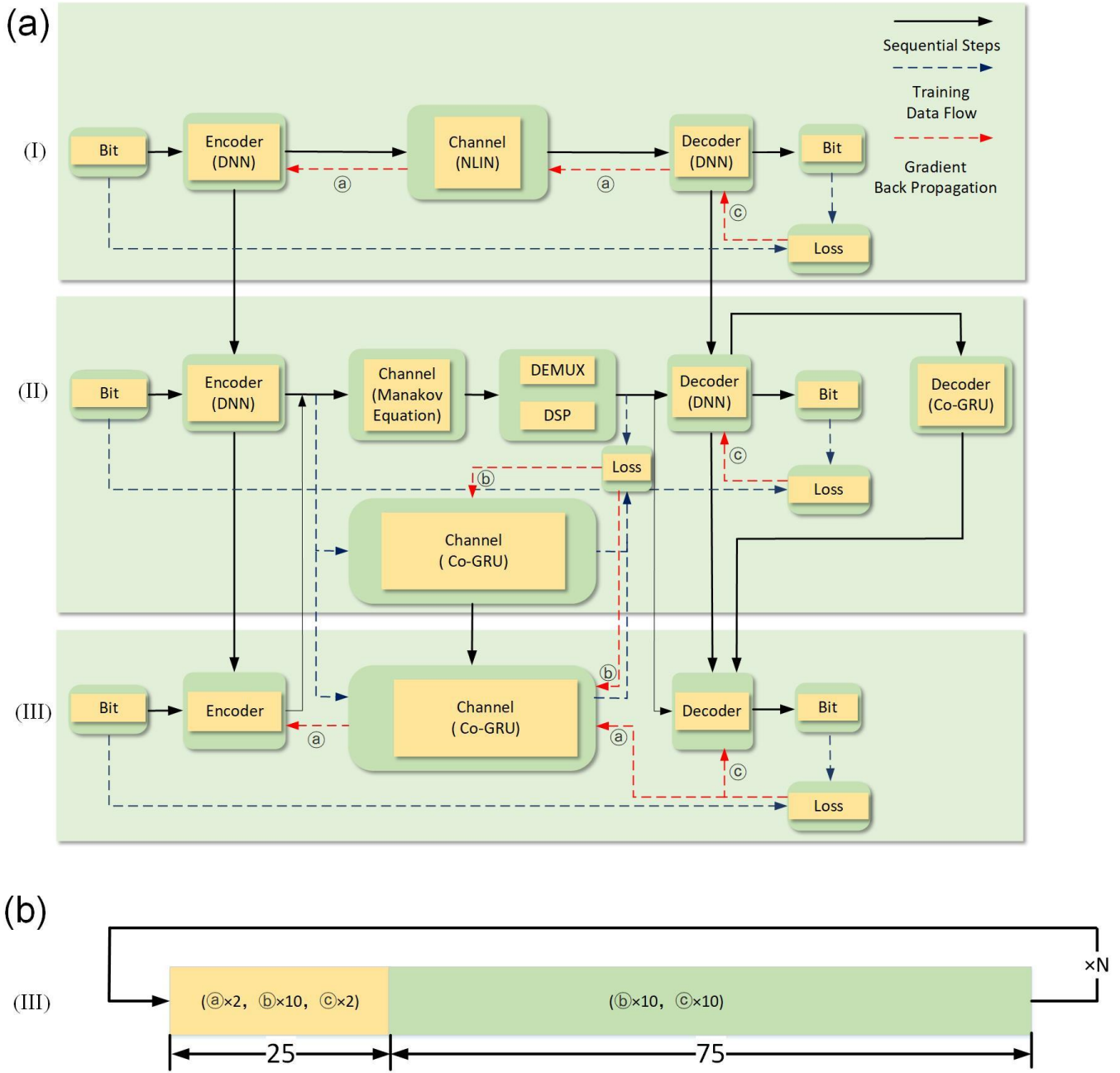

Fig.4. End-to-end training Framework. (a) The E2E training framework consists of Phase I, II & III, where ⓐ, ⓑ & ⓒ respectively represent the operation for training the encoder, channel model and the decoder in different phases. (b)The detailed training process in Phase III.

## IV. SIMULATION AND RESULTS

After E2E training of three Phases, we adopt GMI and $Q^2$-factor as the criterions to compare the performance of DNN and Co-GRU based decoder with regard to conventional QAM systems.

As described above, we simulate a five-channel 64-QAM WDM system. All the results are obtained on Pytorch 1.7.1 with NVIDIA TITAN V Computer Graphics Cards. The E2EDL Hyper-parameters are shown in Table I. The neural network training for channel modeling adopts mean square error (MSE) as the loss function. In addition, for the training of the decoder, the MSE calculated by the decoded bits and the corresponding transmitted bits is used as the loss function, which can accelerate the convergence and maximize the gain of GMI [27].

In order to combat the over-fitting problem, the random number seed for generating the transmitted bits is changed after each training epoch.

Adam optimizer is selected for training, while in Phase I, II & III, the learning rates for ⓐ, ⓑ & ⓒ in Fig. 3 are respectively set as illustrated in Table I.

Table I. The Learning Rate Setting For Training The Encoder (ⓐ), Channel (ⓑ) and the Decoder (ⓒ) in Different Phases

| Phase | ⓐ | ⓑ | ⓒ |
|---|---|---|---|
| I | 0.001 | / | 0.001 |
| II | / | 0.001 | 0.01 |
| III | 0.001 | 0.001 | 0.001 |

### A. Comparison of E2EDL results between NLIN model and Co-GRU model

Fig. 5 compares GMI and $Q^2$-factor through E2E training based on the channel adopting NLIN model in Phase I, with those obtained after further E2EDL training based on the Co-GRU channel model in Phase III. Both the encoder and the decoder are implemented by the DNN. The transmission distance is fixed as 960 km while the launch power is changed from -2.0 to 1.0 dBm. The GMI is estimated through the Gauss-Hermite method introduced in [31], while the $Q^2$-factor is calculated from the BER as Eq. (7).

$$Q^2 = 20\log_{10}(\sqrt{2}erfc^{-1}(2BER)) \quad (7)$$

As shown in Fig.5, both GMI and $Q^2$-factor show an improvement after further E2E training based on the Co-GRU channel model in Phase III compared with that obtained with the NLIN model in Phase I. This validates the difference between the NLIN model and the Co-GRU channel model trained to replace the original channel.

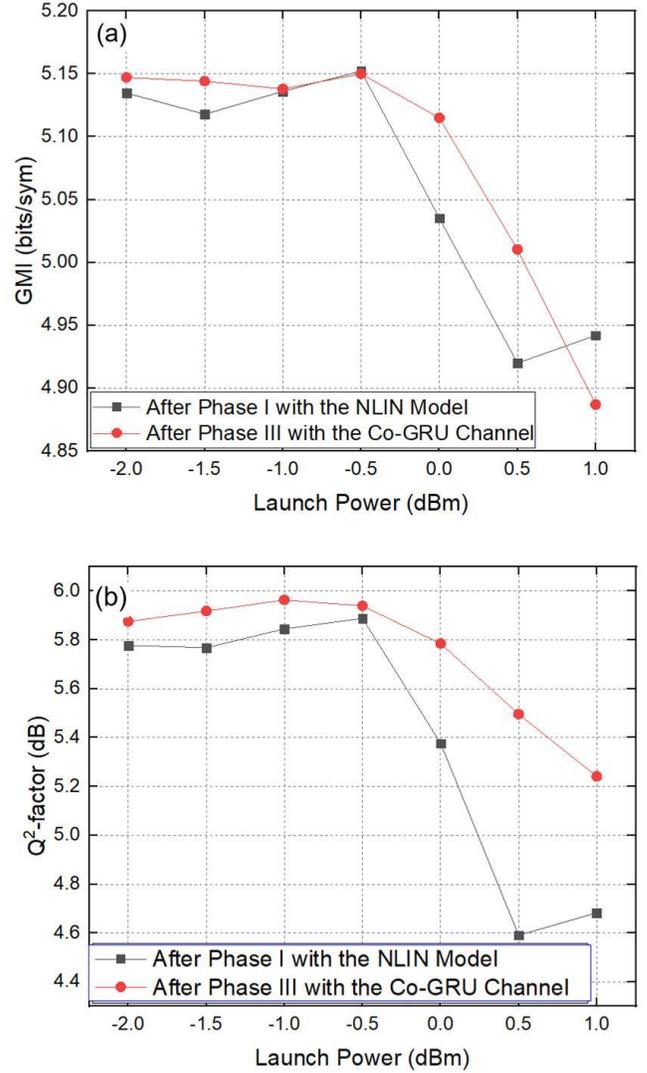

Fig. 5. (a) The GMI and (b) $Q^2$-factor results versus the launch power respectively after E2E trained with the DNN based decoder based on the NLIN model in Phase I and the Co-GRU channel in Phase III. The transmission distance is set as 960km.

### B. The simulation results obtained by respectively adopting DNN and Co-GRU as the decoder

Fig.6 (a) and (b) compare GMI and $Q^2$-factor performance of two types of E2E systems (respectively with the DNN and Co-GRU based decoder) after the three-Phase training described in Fig. 3, with those of the conventional QAM system. It can be seen that for each launch power, the E2E system achieves a GMI and $Q^2$-factor gain compared with the conventional QAM system. Moreover, the E2E system based on Co-GRU decoder obtains a higher GMI and $Q^2$-factor gain, which essentially embodies the equalization ability of the Co-GRU structure for nonlinear distortions.

GMI and $Q^2$-factor gain both increase with launch power at the range of lower than -0.5dBm. As the launch power exceeds 0.0dBm, since NLIN is an amendment to the additive Gaussian noise model, the gap between the NLIN and the SSFM channel distortion enlarges the relative training error in Phase I. Therefore, the E2E compensation ability for nonlinear distortion at large launch power beyond -0.5dBm is limited.

Moreover, around the optimal launch power point, the E2EDL with the Co-GRU decoder achieves a GMI gain up to 0.2bits/sym and a $Q^2$-factor gain up to 0.48dB.

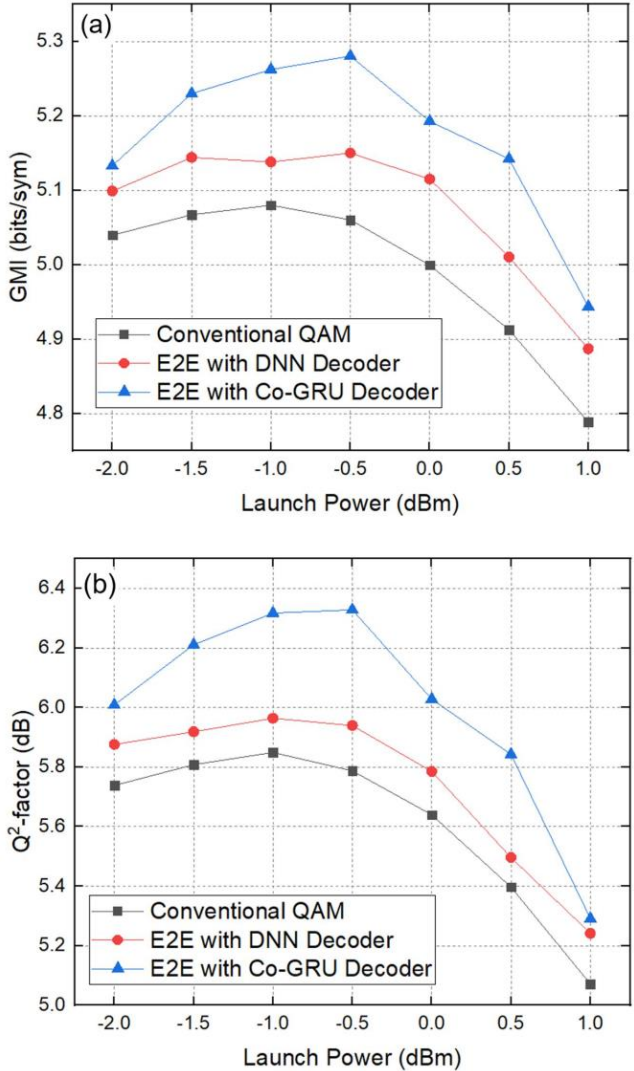

Fig. 6. (a) The GMI and (b) $Q^2$-factor of the conventional QAM system and two types of E2E systems (respectively with the DNN and Co-GRU decoder) after trained with the Co-GRU channel in Phase III, for different launch power. The transmission distance is set as 960km.

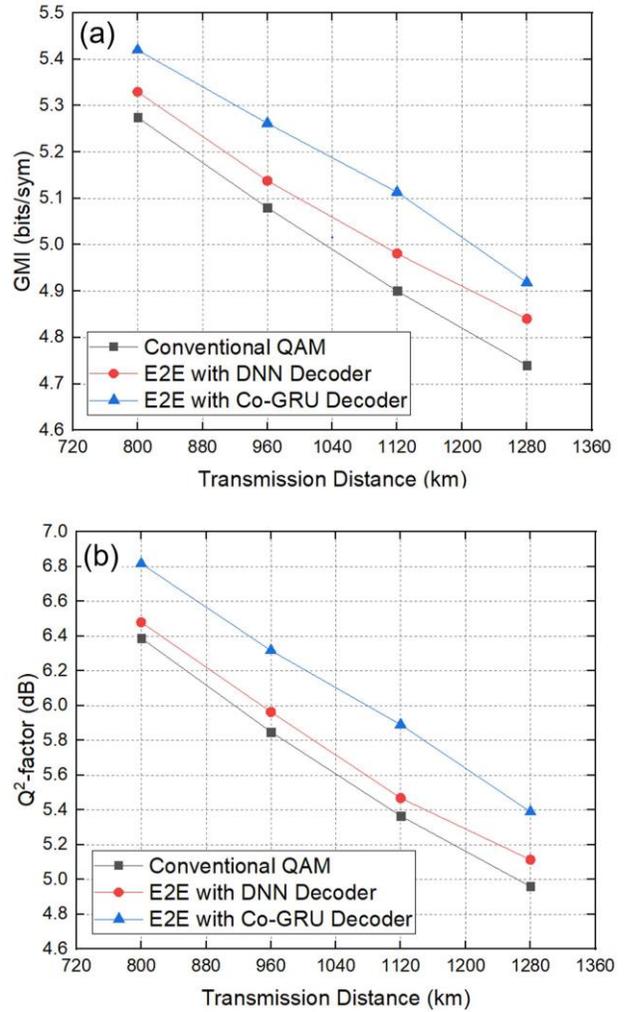

Fig. 7. (a) The GMI and (b) $Q^2$-factor of the conventional QAM system and two types of E2E system (respectively with the DNN and Co-GRU decoder) after trained with the Co-GRU channel in Phase III, versus the transmission distance. The launch power is set as -1.0 dBm.

Fig. 7 shows GMI and $Q^2$-factor for different transmission distance at the launch power of -1.0dBm. The results show that the E2E training effectively compensates for nonlinear distortion along the transmission distance. Moreover, in contrast to the DNN based decoder, the Co-GRU based decoder enables E2EDL to achieve a higher GMI and $Q^2$-factor gain.

Fig. 8 summarizes the constellations obtained respectively after Phase I and III corresponding to each case with different launch power shown in Fig. 6., which shows the adjustment to the constellation exerted by the E2EDL.

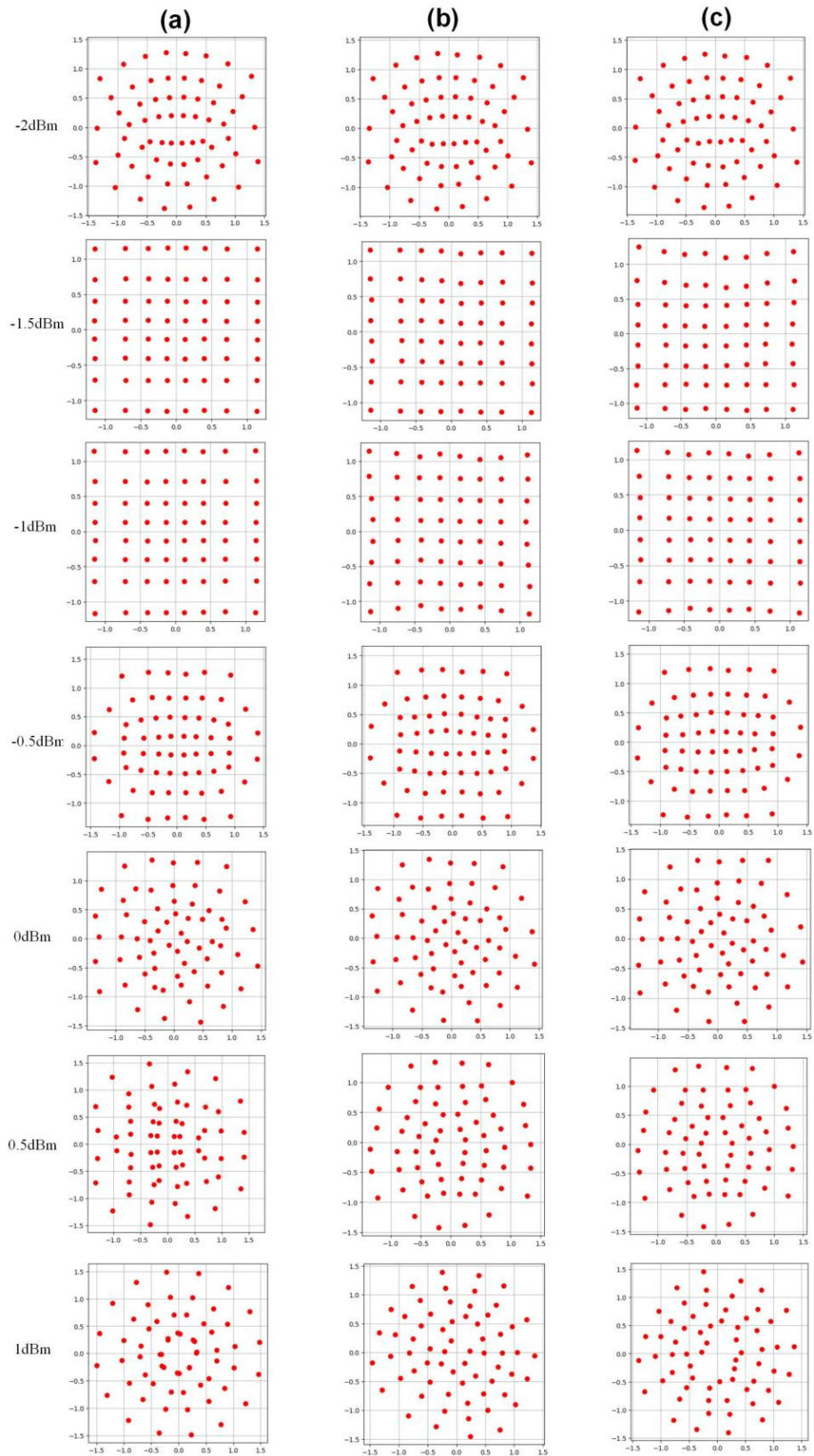

Fig. 8. The constellation of the E2E system corresponding to each case in Fig. 6.
(a) after Phase I; (b) with the DNN decoder after Phase III; (c) with the Co-GRU decoder after Phase III.

*C. Loss trend during E2E training*

Fig.9. shows the MSE loss during training the Co-GRU channel model and the BER during training the decoder in Phase II. The launch power is set as -1.0 dBm and the transmission distance is set as 960km.

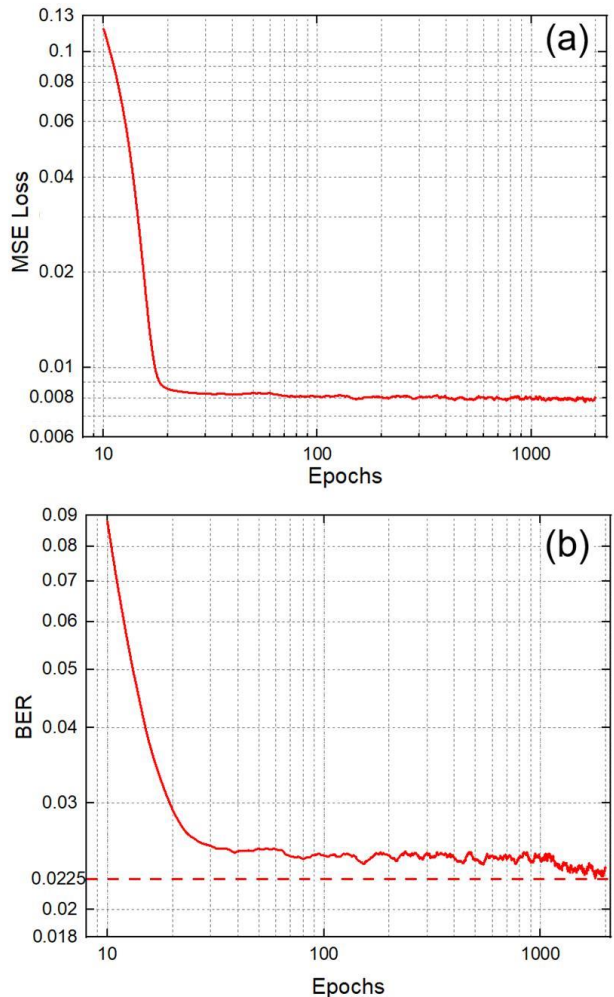

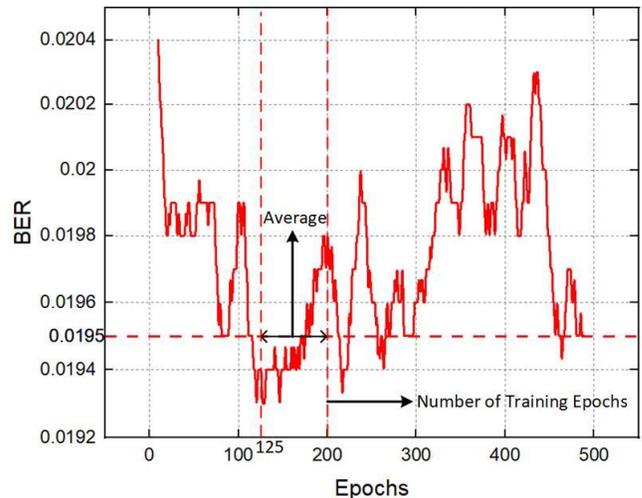

Fig. 10. The BER of different launch power after further E2E training in Phase III with the transmission distance as 960km and launch power of -1.0 dBm;

Fig. 9. (a) The MSE loss and (b) the BER of training the Co-GRU channel model and decoder in Phase II with the transmission distance as 960km and the launch power as -1.0dBm.

As shown in Fig.9(a) and (b), for training the Co-GRU channel and decoder, the MSE loss and the BER decreases rapidly while converging to the minimum in less than 100 Epochs, which demonstrates the effectiveness of the Co-GRU structure for both the channel modeling and the decoder implementation. Meanwhile, both the Co-GRU channel and decoder are trained 2000 epochs to minimize the MSE loss and BER. Furthermore, for training the Co-GRU decoder, the BER is minimized to about 0.0225 after Phase II as shown in Fig.9 (b).

It should be noted that there exists a gap between the original channel and the Co-GRU channel model. Moreover, the decoder is trained based on the SSFM channel while the encoder is updated through the Co-GRU channel model using the gradient BP method. Due to the above reason, in order to approach the optimum through E2E training, the alternate training method is adopted in Phase III. That is, for each 100 training epochs periodically, while the encoder and the decoder joint training is implemented in the first 25 epochs, only the decoder is trained in the last 75 training epochs.

The BER through the training in Phase III is shown in Fig.10 with launch power of -1.0dBm as an example. For each period containing 100 training epochs, the average BER is calculated from the last 75 epochs. The final epoch of the period with the lowest average BER, is determined as the epoch to stop training. Due to that the average BER from $125^{th}$ to $200^{th}$ training epochs has the lowest value as 0.01950, the total number of the training epochs is selected artificially as 200. Comparing Fig. 10 and Fig.9 (b), it is shown that the BER is reduced from 0.0225 after Phase II to 0.0195 through further E2E training in Phase III.

V. CONCLUSION

In this work, we first discuss the modeling of the optical fiber communication system through the Co-GRU neural network structure and the simplified calculation scheme. Through Co-GRU, we can accurately model both linear and nonlinear interference in optical fiber transmission, and quickly calculate gradient back-propagation. The E2E framework proposed consists of 3 training Phases. The encoder and the decoder are trained firstly through the NLIN model in Phase I, and further trained on the Co-GRU channel in Phase II and III. We compare the E2E training results through the NLIN model with that obtained through further training on the Co-GRU channel model. Implementing the decoder-only training following the encoder and the decoder joint training periodically, the proposed E2E scheme can combat the difficulty in arriving at the optimum through gradient BP, which results from the gap between the Co-GRU channel and the SSFM channel. Moreover, for 32G Baud 960km 5 channel dual-polarization WDM transmission, the E2E system with the Co-GRU based decoder achieves the $Q^2$-factor gain up to 0.48dB and the GMI gain up to 0.2bits/sym compared with the conventional 64QAM system at around the optimal launch power point. The above results indicate the application feasibility of the Co-GRU structure for both the channel modeling and the decoder performance improvement in the E2EDL framework design.